\documentclass[3p]{elsarticle}

\usepackage{amssymb}
\usepackage{amsmath}
\usepackage{url}
\usepackage{lineno}
\usepackage[dvipsnames]{xcolor}

\journal{Applied Mathematics and Computation}

\begin{document}

\begin{frontmatter}

\title{Context-sensitive norm enforcement reduces sanctioning costs in spatial public goods games}

\author[label1]{Hsuan-Wei Lee\corref{mail1}}
\cortext[mail1]{Corresponding authors. Email addresses: hsl324@lehigh.edu; szolnoki.attila@ek-cer.hu}
\author[label2]{Colin Cleveland}
\author[label3]{and Attila Szolnoki\corref{mail1}}

\address[label1]{College of Health, Lehigh University, Bethlehem, PA 18015, USA}
\address[label2]{Department of Informatics, King's College London, London, UK}
\address[label3]{Institute of Technical Physics and Materials Science, Centre for Energy Research, P.O. Box 49, H-1525 Budapest, Hungary}

\begin{abstract}
Uniform punishment policies can sustain cooperation in social dilemmas but impose severe costs on enforcers, creating a second-order free-rider problem that undermines the very mechanism designed to prevent exploitation. We show that the remedy is not a harsher stick but a smarter one. In a four-strategy spatial public-goods game we pit conventional punishers, who levy a fixed fine, against norm-responsive punishers that double both fine and cost only when at least half of their current group already cooperates. Extensive large scale Monte Carlo simulations on lattices demonstrate that context-sensitive punishment achieves complete defector elimination at fine levels 15\% lower than uniform enforcement, despite identical marginal costs per sanctioning event. The efficiency gain emerges because norm-responsive punishers conserve resources in defector-dominated regions while concentrating intensified sanctions at cooperative-defector boundaries, creating self-reinforcing fronts that amplify the spread of prosocial behavior. These findings reveal that enforcement efficiency can be dramatically improved by targeting punishment at cooperative-defector interfaces rather than applying uniform sanctions, offering quantitative guidelines for designing adaptive regulatory mechanisms that maximize compliance while minimizing institutional costs.
\end{abstract}


\end{frontmatter}

\section{Introduction}
\label{sec:intro}
Cooperation flourishes when individuals are willing to incur short-term costs for the collective good, yet in virtually every social dilemma the immediate temptation is to free-ride~\cite{sigmund1999evolutionary, perc2010coevolutionary}. This fundamental tension between individual rationality and collective welfare has inspired decades of research into mechanisms that can tip the balance toward prosocial behavior~\cite{ostrom2000collective, kirman2010complex, szolnoki2013information, tilman2019localized}. Costly peer punishment can suppress this temptation by raising the expected price of defection, but uniform sanctioning policies impose severe fiscal burdens on enforcers while often failing to maximize deterrent effects~\cite{baumard2010has, szolnoki2011competition, perc2015double}. Alternative enforcement architectures, such as global exclusion mechanisms implemented by centralized institutions, can address both first- and second-order free-riding problems through different institutional pathways~\cite{wang2022replicator}. Unless the burden is reduced or redistributed, punishers confront a second-order dilemma in which they themselves are exploited by pure cooperators who refuse to contribute to this incentive, but enjoy its positive consequences. How to sustain an effective deterrent without exhausting the enforcers therefore remains a central puzzle for evolutionary biology~\cite{wakano2007evolution, santos2011evolution, raihani2012punishment}, behavioral economics~\cite{egas2008economics, hetzer2013co, angelovski2018behavioral} and public-policy design~\cite{henrich2006cooperation, isakov2012evolution, vasconcelos2022punishment}.

Everyday practice hints at a pragmatic answer. Across legal, organizational and informal settings sanctions are rarely uniform: auditors prioritize firms that are otherwise compliant, traffic police patrol safe districts more aggressively than crime-ridden ones, neighborhoods ostracize the rare cheat far more harshly than the habitual free-rider next door, and even schoolchildren modulate their scolding to the behavior of the majority. These observations suggest that effective enforcement systems naturally evolve toward context-dependent strategies that concentrate limited resources where they can achieve maximum behavioral change~\cite{downs1997enforcement, frey2012evolutionary, botta2024discipline, gao2025evolutionary}. Institutional frameworks combining monitoring and reporting mechanisms have demonstrated how centralized governance can efficiently solve collective-risk dilemmas by strategically targeting enforcement based on detected violations~\cite{he2019central}. Such patterns suggest that coupling punishment to local conditions can conserve resources while keeping deterrence intact \cite{perc_njp12,chen_njp14}. Empirical studies of punishment in daily life confirm that humans employ context-sensitive sanctioning strategies, adjusting between direct confrontation and indirect punishment (gossip, exclusion) based on factors such as violation severity, power differentials, and relationship value~\cite{molho2020direct}, demonstrating that conditional enforcement is a fundamental feature of human social behavior. Threshold-based switching between exclusion and reward mechanisms further illustrates how moderate conditional responses can establish high cooperation levels more effectively than uniform institutional policies~\cite{he2023advantage}.

Experimental evidence supports the existence of conditional punishment strategies in human populations, where individuals adjust their sanctioning behavior based on prevailing cooperation levels (norm enforcement) or punishment patterns displayed by others (conformist punishment)~\cite{li2021descriptive}, suggesting that context-dependent enforcement may reflect fundamental behavioral tendencies rather than mere theoretical constructs. Recent advances in spatial evolutionary games have explored various sophisticated mechanisms for sustaining cooperation, including mercenary punishment systems~\cite{lee2022mercenary}, taxation-based enforcement~\cite{lee2024supporting}, asymmetric punishment in interdependent networks~\cite{guo2024evolution}, and adaptive network structures~\cite{lee2018evolutionary, yang2019evolution, lee2025enhancing}, as well as punishment strategies with adaptive feedback mechanisms~\cite{hua2023facilitating} and state-feedback systems that adjust punishment intensity based on collective contribution levels~\cite{wang2025evolutionary}. Additionally, dynamic interaction models incorporating queueing systems and reputation mechanisms have demonstrated how temporal coordination and past behavior influence cooperative outcomes~\cite{zhang2025spatial}. While these studies demonstrate the potential for dynamic punishment mechanisms, yet the potential for context-dependent punishment intensity that responds specifically to local cooperation density remains largely unexplored.

Spatial public goods games provide a natural laboratory for this inquiry because network reciprocity already lets cooperators form protective islands and punishment can reinforce these clusters \cite{szolnoki2009topology, helbing2010punish, xia2012effects, yang2015mutual, chen2015competition, quan2023cooperation, flores2024evolution}. The critical question is what happens when fines intensify only where surrounding cooperation is high. The key insight we explore is that punishment intensity should scale with local cooperation density rather than being applied uniformly across all social contexts. We address the question with a four-strategy model comprising cooperators $C$, defectors $D$, conventional punishers $P_{1}$ who always impose a baseline fine $\beta$, and norm-responsive punishers $P_{2}$ who double both fine and cost whenever at least half of their current group members cooperate. Crucially, a $P_{2}$ agent makes this assessment independently in each of the five overlapping groups to which it belongs on a lattice, so the same individual can punish hard at one interface, softly in a second, and not at all in a third. This built-in flexibility sets the stage for an evolutionary contest between blunt and nuanced enforcement and allows us to test whether smarter really is better than stronger. This approach builds on previous work demonstrating how intelligent strategies can reshape social dilemma outcomes \cite{macy2002learning, bogaert2008social, lee2024suppressing, mintz2025evolutionary} and how selective mechanisms can achieve disproportionate welfare improvements \cite{su2017evolutionary, sendina2020diverse, szolnoki2020blocking, engel2021managing, lee2021small, sun2023state}.

Extensive Monte Carlo simulations on lattices up to $1200\times1200$ reveal a striking pattern. Although $P_{1}$ and $P_{2}$ incur identical marginal costs per sanctioning event, the context-aware strategy achieves complete defector elimination at fine levels approximately 15\% lower than required by uniform punishment policies. The mechanism is microscopically clear: $P_{2}$ conserves its budget inside defector strongholds and unleashes its double fine only along cooperative fronts, where each newly converted neighbor widens the majority that justifies escalation. Model 2, which escalates in defector territory, squanders resources and collapses, leaving the field to either $P_{1}$ or unpunished free-riding. The same qualitative outcome holds across a broad span of synergy factors, punishment costs, initial conditions and weak mutation rates.

These results advance the theory of altruistic enforcement and offer a blueprint for policy. Smart punishment need not exceed blunt punishment in severity; it must simply target the brink. By aligning sanction intensity with local compliance, institutions can achieve higher cooperation at lower social cost. Beyond advancing evolutionary game theory, these findings provide quantitative guidelines for designing adaptive enforcement mechanisms in domains ranging from tax compliance to environmental regulation. The remainder of the paper is organized as follows. Section \ref{sec:methods} defines the model and outlines simulation techniques, Section \ref{sec:results} presents phase diagrams and microscopic evidence, and Section \ref{sec:conclusions} summarizes key insights and future directions.

\section{Methods}
\label{sec:methods}

\subsection{Four-Strategy Competition Model}

We implement a spatial evolutionary game on an $L \times L$ square lattice with periodic boundary conditions, where each node hosts a single player engaged in overlapping public goods interactions with its four nearest neighbors. Our model features four competing strategies that coevolve in a spatial environment. According to the traditional public goods game setup, neighbors form groups of $N=5$ members, a focal player and its four nearest neighbors, and decide simultaneously whether to contribute to a common pool or not. Cooperators ($C$) contribute $c=1$ to the public pool but do not engage in punishment activities. Defectors ($D$) contribute nothing, thereby exploiting others' contributions. Conventional punishers ($P_1$) contribute $c=1$ and impose a fixed fine $\beta$ on defectors at cost $\gamma$ per punishment act. Norm-responsive punishers ($P_2$) contribute $c=1$ and modulate punishment intensity based on local cooperation levels. This per-group assessment equips every $P_{2}$ player with an intrinsic sensor that throttles enforcement to the immediate social temperature, turning punishment into a locally adaptive control system.

The crucial innovation lies in making punishment intensity contingent on local social context rather than applying uniform sanctions across all environments. The strategic distinction between $P_1$ and $P_2$ lies in their sanctioning approach. While $P_1$ applies uniform punishment regardless of social context, $P_2$ assesses the cooperation level $\phi = (N_C + N_{P_1} + N_{P_2})/N$ within each group independently, where $N=5$ is the group size. Critically, a single $P_2$ player may simultaneously enforce different punishment intensities across different groups—applying standard sanctions in low-cooperation environments while intensifying punishment in cooperative contexts. This group-specific response allows $P_2$ players to adapt their sanctioning strategy to local conditions, potentially conserving resources in defector-dominated groups while reinforcing cooperation where it has already gained a foothold.

We investigate two complementary variants of norm-responsive punishment. In Model 1, $P_2$ intensifies punishment when $\phi \geq 1/2$ (cooperative environments), doubling both fine and related cost in groups where cooperators form the majority. In Model 2, the logic is reversed: $P_2$ intensifies punishment when $\phi < 1/2$ (defector-dominated environments), applying stronger sanctions precisely where cooperation is struggling to establish itself. Both models maintain identical fine-to-cost ratios ($\beta/\gamma$) for both punisher types across all conditions, eliminating trivial efficiency advantages and isolating the effects of context-sensitivity.

\subsection{General Payoff Formalism}

The public goods interaction proceeds in sequential phases: first contributions are collected, then multiplied by the synergy factor and redistributed, and finally punishment is applied. Let $N_X$ denote the number of players with strategy $X \in \{C, D, P_1, P_2\}$ in a focal group, and $\phi = (N_C + N_{P_1} + N_{P_2})/N$ represent the local cooperation level. The payoff structure switches discretely at the cooperation threshold $\phi = 1/2$, creating distinct behavioral regimes that drive the emergence of spatial patterns.

For Model~1, where $P_2$ intensifies punishment in cooperative environments ($\phi \geq 1/2$), the payoffs in low-cooperation regimes ($\phi < 1/2$) are given by:
\begin{align}
	\Pi_C &= r\frac{N_C + N_{P_1} + N_{P_2}}{N} - 1\\
	\Pi_D &= r\frac{N_C + N_{P_1} + N_{P_2}}{N} - \beta(N_{P_1} + N_{P_2})\\
	\Pi_{P_1} &= \Pi_C - \gamma N_D\\
	\Pi_{P_2} &= \Pi_C - \gamma N_D\,.
\end{align}

When cooperation dominates ($\phi \geq 1/2$), the payoffs transform to:
\begin{align}
	\Pi_C &= r\frac{N_C + N_{P_1} + N_{P_2}}{N} - 1\\
	\Pi_D &= r\frac{N_C + N_{P_1} + N_{P_2}}{N} - \beta(N_{P_1} + 2N_{P_2})\\
	\Pi_{P_1} &= \Pi_C - \gamma N_D\\
	\Pi_{P_2} &= \Pi_C - 2\gamma N_D\,.
\end{align}

For Model~2, where $P_2$ intensifies punishment in defection-dominated groups, the payoff structure is inverted. When defection prevails ($\phi < 1/2$):
\begin{align}
	\Pi_C &= r\frac{N_C + N_{P_1} + N_{P_2}}{N} - 1\\
	\Pi_D &= r\frac{N_C + N_{P_1} + N_{P_2}}{N} - \beta(N_{P_1} + 2N_{P_2})\\
	\Pi_{P_1} &= \Pi_C - \gamma N_D\\
	\Pi_{P_2} &= \Pi_C - 2\gamma N_D\,.
\end{align}

And, when cooperation dominates ($\phi \geq 1/2$):
\begin{align}
	\Pi_C &= r\frac{N_C + N_{P_1} + N_{P_2}}{N} - 1\\
	\Pi_D &= r\frac{N_C + N_{P_1} + N_{P_2}}{N} - \beta(N_{P_1} + N_{P_2})\\
	\Pi_{P_1} &= \Pi_C - \gamma N_D\\
	\Pi_{P_2} &= \Pi_C - \gamma N_D\,.
\end{align}

The total payoff for each player is calculated by summing their earnings across all five groups in which they participate. This mechanism captures the cumulative effect of context-dependent strategies operating simultaneously in multiple overlapping social environments, a critical feature that enables the complex spatial patterns observed in our simulations.

\subsection{Evolutionary Dynamics}

Strategy evolution proceeds through asynchronous imitation with stochastic noise. In each elementary update, a randomly selected player $i$ and one of its von~Neumann neighbors $j$ are compared. Player $i$ adopts player $j$'s strategy with probability $W_{i \leftarrow j} = [1 + \exp{((\Pi_i - \Pi_j)/K)}]^{-1}$, where $K = 0.1$ controls selection intensity, balancing between deterministic optimization ($K \rightarrow 0$) and neutral drift ($K \rightarrow \infty$). The finite noise parameter K enables occasional irrational strategy choices, helping the system avoid local equilibria and ensuring robust exploration of the strategy space. Extensive testing across $K \in [0.05, 0.5]$ confirms that while absolute critical values may shift quantitatively, the qualitative advantage of context-sensitive punishment over uniform enforcement remains invariant, consistent with established results for spatial evolutionary games~\cite{szabo2005phase, vukov2006cooperation}. To prevent trapping in absorbing states and ensure ergodic exploration of the strategy space, we implement rare point mutations: with probability $\mu = 10^{-6}$ per update, a player randomly switches to one of the three alternative strategies. The selection intensity $K = 0.1$ ensures realistic levels of behavioral noise while maintaining sufficient selective pressure, and the mutation rate $\mu = 10^{-6}$ provides ergodic exploration without overwhelming evolutionary forces. We investigate evolutionary dynamics on an $L \times L$ square lattice with periodic boundary conditions, using system sizes $L \in \{200, 400, 600, 1200\}$ to eliminate finite-size effects. 

Each lattice site hosts a single player who participates in $N=5$ overlapping public goods games: one centered on itself and four where it serves as a member in its von Neumann neighbors' groups. This spatial structure facilitates the emergence of strategy domains and complex interfaces crucial for understanding context-dependent punishment. One Monte Carlo step (MCS) consists of $N = L^2$ elementary updates, giving each player one expected opportunity to revise its strategy. Simulations run for at least $10^5$ MCS, with initial $2 \times 10^4$ steps discarded as transient before computing equilibrium distributions averaged across 30 independent realizations. This extensive sampling ensures statistical reliability of our results, particularly near critical transition points where fluctuations can be substantial.

To investigate initialization effects, we implement three distinct protocols: random uniform distribution with equal probability assignment for each strategy, stripe configuration with four horizontal bands of pure strategies, and patchy structure comprising $20 \times 20$ blocks of homogeneous strategy assignments. Ultra-large scale simulations ($L = 1200$) at fixed $(r, \gamma) = (2.5, 0.4)$ with $\beta \in [0.4, 0.6]$ reveal surprising initialization dependencies. Across 30 independent trials of $10^5$ MCS each, patchy initializations consistently produce stationary distributions that differ quantitatively from those achieved through random initial conditions, even after extensive relaxation periods. The well-defined interfaces established in patchy initializations appear to stabilize certain coexistence patterns that would otherwise be eliminated through stochastic fluctuations in random starts. 

For methodological robustness, we determined that random initialization with weak mutation ($\mu = 10^{-6}$) provides the most reliable convergence to true equilibrium distributions, enabling accurate mapping of phase boundaries. This approach optimally balances computational efficiency with ergodic exploration of the strategy space. This initialization dependence reveals the path-dependent nature of spatial evolutionary dynamics and demonstrates that our random initialization protocol with weak mutation provides the most reliable method for mapping true equilibrium phase boundaries.

\section{Results}
\label{sec:results}

\subsection{Global Phase Structure on The Punishment Plane}
\label{subsec:phase}

\begin{figure}[h]
	\centering
	\includegraphics[width=0.9\textwidth]{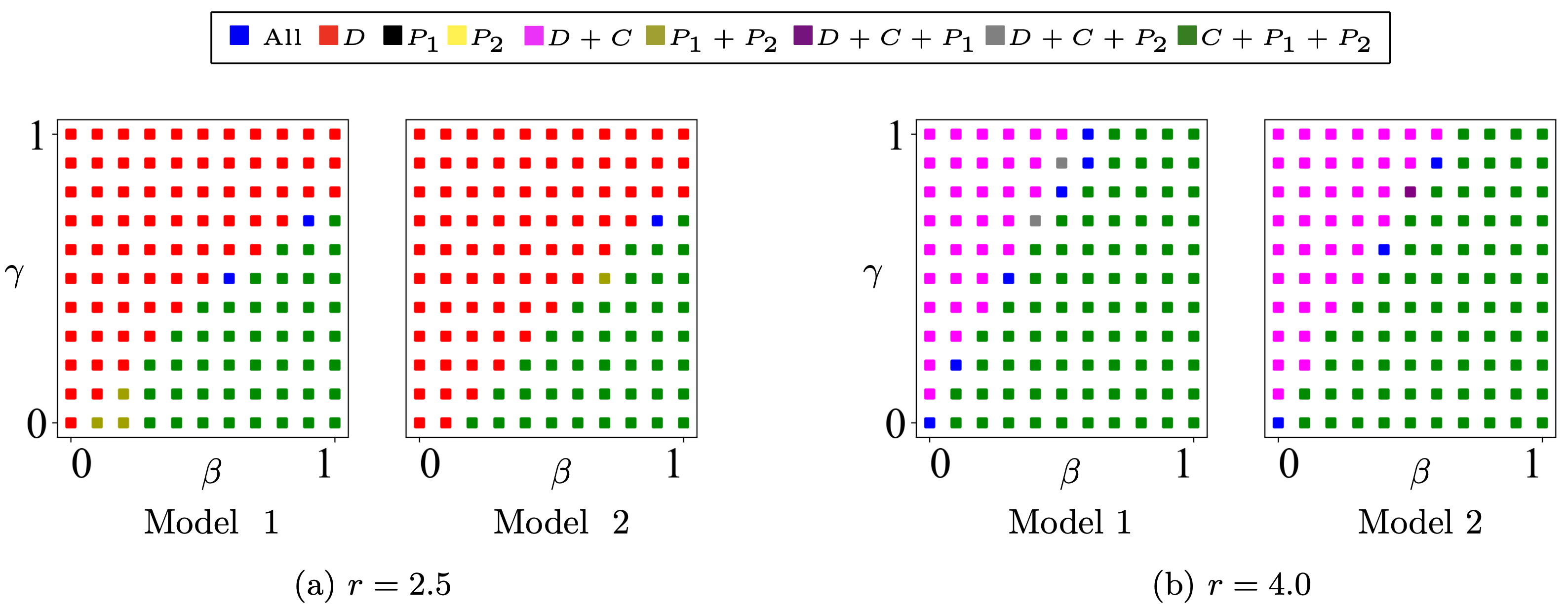}
\caption{Stationary strategy landscapes in the $\beta$–$\gamma$ plane. Panel~(a) $r=2.5$ and panel~(b) $r=4.0$. Left squares: Model~1 (escalation in cooperative groups); right squares: Model~2 (escalation in uncooperative groups). Each colored square represents the set of strategies that coexist in the stationary state reached after $10^{6}$\, MCS on a $600\times600$ lattice; colors are decoded in the legend. The phase diagrams reveal that Model 1 consistently achieves defector extinction at fine levels 15\% lower than Model 2 across the entire feasible parameter range, demonstrating the superior efficiency of context-sensitive punishment that targets cooperative-defector interfaces rather than applying uniform sanctions.}
	\label{fig:phase}
\end{figure}
Systematic exploration of the punishment parameter space reveals the fundamental trade-offs between enforcement costs and deterrence effectiveness, exposing distinct regimes where context-sensitive strategies achieve superior performance. Figure~\ref{fig:phase} summarizes the evolutionary outcomes of the four-strategy public-goods game for every tested pair of punishment parameters. The fine $\beta$ and the cost $\gamma$ are sampled in steps of $0.05$, so that each colored square represents the long-run state reached from a random initial condition at that parameter combination. Two representative synergy factors are shown: panel~(a) uses $r=2.5$, lying below the threshold where network reciprocity can protect cooperation unaided, whereas panel~(b) uses $r=4.0$, well above that threshold. Within each panel the left column corresponds to Model~1, where a $P_{2}$ player doubles both fine and cost when its own group is at least half cooperative; the right matrix shows Model~2, in which the same escalation is triggered when cooperation is below one half. Each square shows the set of strategies that coexist in the stationary state reached after $10^6$ MCS on a $600\times600$ lattice. The legend placed above the plots lists the nine possible destinations ranging from pure defection to the full four-strategy coexistence. The phase diagrams reveal that Model~1 consistently achieves defector extinction at fine levels 15\% lower than Model~2 across the entire feasible parameter range.

For the lower synergy $r=2.5$ (Figure~\ref{fig:phase} , panel~(a)) cooperation collapses unless punishment intervenes. The diagrams are therefore divided by a critical frontier $\beta=\beta_{\mathrm c}(\gamma)$. Below the critical threshold $\beta_c(\gamma)$ defectors dominate globally (red), while above this boundary punishment becomes sufficiently effective to establish cooperative equilibria dominated by punisher strategies (light green). In Model 1 the frontier sits noticeably lower than in Model~2, revealing that context-sensitive escalation in already cooperative groups is more efficient design. This efficiency advantage stems from Model~1's ability to concentrate punishment resources at cooperative-defector interfaces where conversion probability is maximized, rather than dispersing effort throughout defector-dominated regions. The superiority persists until the cost reaches $\gamma\simeq0.7$, beyond which both models lose the ability to fund sanctions and the population reverts to defection. When the synergy increases to $r=4.0$ (Figure~\ref{fig:phase} , panel~(b)), a band of $C$-$D$ coexistence (magenta) appears at small fines, showing that network reciprocity by itself can hold a mixed state. Once $\beta$ exceeds the coexistence window the contrast between the two designs reappears: Model~1 supports an expansive domain where norm-responsive punishers sometimes dominate (gray), whereas Model~2 is governed almost entirely by conventional $P_{1}$ players (purple). All transition lines are shifted leftwards relative to $r=2.5$, reflecting the reduced selective pressure for punishment when cooperation already enjoys partial support.

Across the entire parameter plane the minimum fine required to eradicate defectors is $10$-$20\%$ lower for Model~1 than for Model~2, even though the fine-to-cost ratio is identical. The systematic saving underscores a general design rule: punishment that escalates where cooperation is already strong leverages resources far more effectively than escalation in hostile territory. This insight points to concrete guidelines for crafting economical yet potent enforcement mechanisms in human and artificial societies.

\subsection{Critical Fine Thresholds}
\label{subsec:beta_scan}
\begin{figure}[h!]
	\centering
	\includegraphics[width=0.88\textwidth]{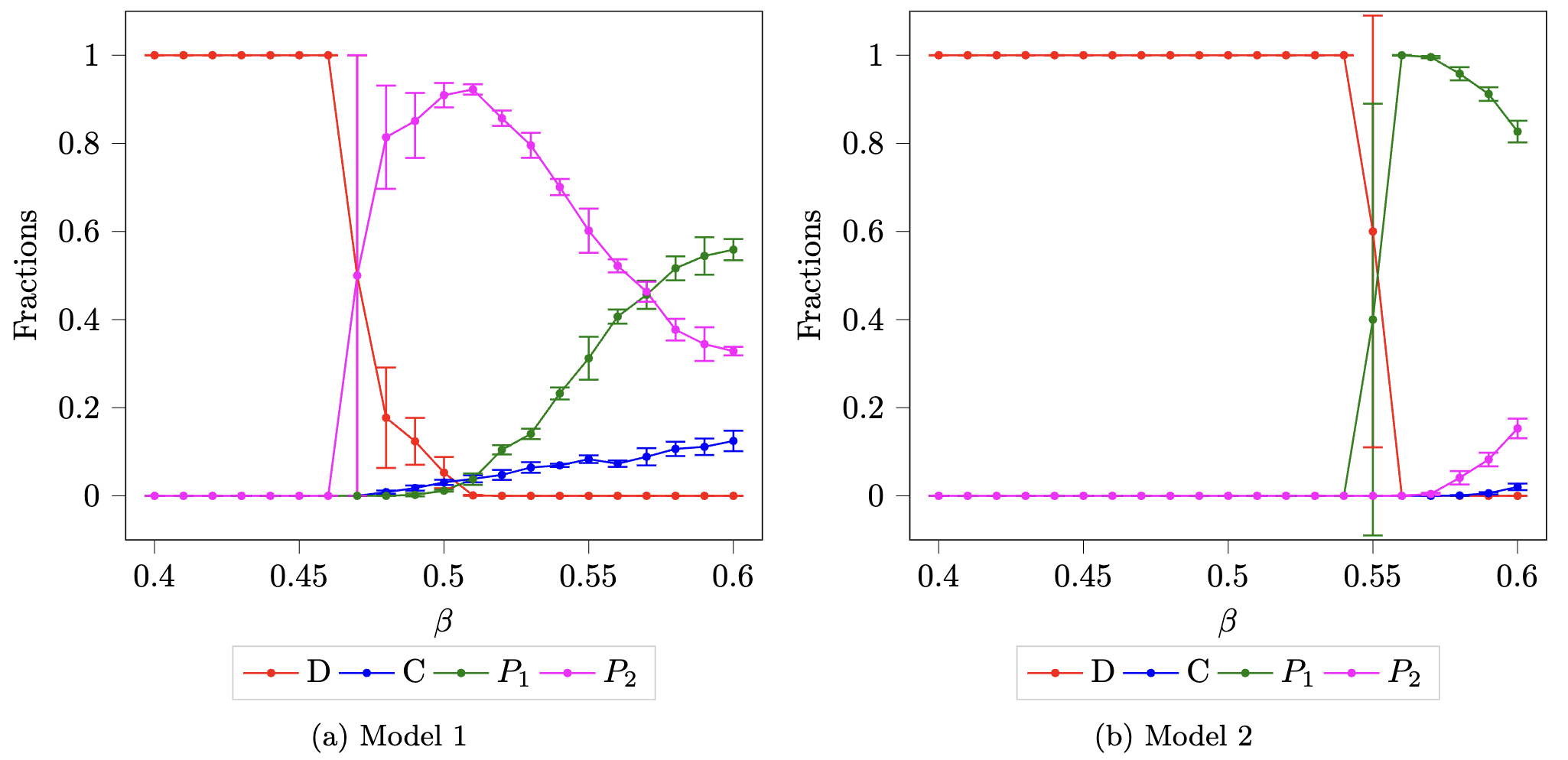}
\caption{High-resolution sweep of the stationary fractions of defectors ($D$), cooperators ($C$), conventional punishers ($P_{1}$), and norm-responsive punishers ($P_{2}$) as the fine $\beta$ is varied from 0.4 to 0.6 at fixed $r=2.5$ and $\gamma=0.4$. Panel~(a) Model~1, escalation in cooperative groups; panel~(b) Model~2, escalation in uncooperative groups. Vertical error bars denote one standard deviation across 30 realizations. The critical fines at which defectors lose stability are $\beta_{\mathrm c}^{(1)}\simeq0.46$ and $\beta_{\mathrm c}^{(2)}\simeq0.54$, quantifying the efficiency advantage of context-aware punishment. This  demonstrates that targeting punishment intensity based on local cooperation density achieves superior enforcement efficiency, with Model 1 showing explosive growth of norm-responsive punishers $P_2$ immediately above the critical threshold, while Model 2 relies primarily on conventional punishers $P_1$ and requires higher fines to achieve equivalent defector suppression.}

	\label{fig:beta_scan}
\end{figure}

Figure~\ref{fig:beta_scan} dissects the phase transition along a single cut of parameter space, keeping the synergy factor and punishment cost fixed at $(r,\gamma)=(2.5,0.4)$ while increasing the fine $\beta$ in steps of $\Delta\beta=0.01$ from 0.4 to 0.6. Each point is the average of 30 independent realizations on a $1200\times1200$ lattice evolved for $10^{5}$\, MCS; error bars show one standard deviation. The simulation duration of $10^5$ MCS is sufficient to reach dynamical equilibrium, as verified by monitoring strategy fraction convergence and ensuring that fluctuations around mean values remain within one standard deviation for the final $2 \times 10^4$ steps.

Model~1 (Figure~\ref{fig:beta_scan}, panel~(a)) in which $P_{2}$ escalates sanctions inside cooperative groups, displays an abrupt transition at $\beta_{\mathrm c}^{(1)}\simeq0.46$. Below $\beta_c^{(1)}$ defectors maintain complete dominance; immediately above this threshold norm-responsive punishers undergo explosive growth from negligible levels to 80\% abundance, simultaneously driving defector prevalence below 20\% through a sharp first-order-like transition. As $\beta$ continues to rise the $P_{2}$ fraction peaks near $0.51$ then declines smoothly while conventional punishers $P_{1}$ and residual cooperators grow at their expense. Throughout the entire super-critical range the share of $P_{1}$ never overtakes $P_{2}$, confirming that context-sensitive escalation is selectively favored even though the two punisher types bear the same fine-to-cost ratio in low-cooperation groups. The sustained dominance of $P_2$ over $P_1$ throughout the supercritical regime demonstrates that context-awareness provides a fundamental evolutionary advantage beyond mere parameter tuning. Model~2 (Figure~\ref{fig:beta_scan}, panel~(b)) tells the mirror story. Because escalation is now triggered in hostile territory the critical fine is pushed to $\beta_{\mathrm c}^{(2)}\simeq0.54$, a full $((0.54-0.46)/0.54)\approx 15\%$ higher than in Model~1. At the transition $P_{1}$, not $P_{2}$, becomes the dominant strategy and rapidly excludes defectors; the norm-responsive type survives only at the percent level for the largest fines examined. The relative inefficiency of Model~2 is thus visible in two complementary metrics: a higher $\beta_{\mathrm c}$ and the absence of $P_{2}$ dominance once cooperation has been restored.

These differences are robust to changes in lattice size ($600\le L\le1200$) and decision noise ($0.05\le K\le0.5$). Taken together they demonstrate that where punishment is intensified matters at least as much as how much is spent. Targeting scarce resources to already cooperative pockets reduces the fine necessary to eradicate defection, offering a quantitative benchmark for the design of adaptive sanctioning schemes.

\subsection{Influence of Starting Heterogeneity and Weak Mutation}
\label{subsec:init}

The path-dependent nature of spatial evolutionary dynamics raises the crucial question of whether observed macroscopic patterns represent genuine equilibrium properties or artifacts of specific initialization protocols. We therefore repeated the key sweep at \((r,\gamma)=(2.5,0.4)\) on an ultra-large \(1200\times1200\) system, where finite-size artifacts are negligible, and monitored how the stationary composition depends on both the initial spatial pattern and the presence of rare mutations. Three complementary protocols were explored. (i) A maximally disordered start in which every lattice site is assigned \(D\), \(C\), \(P_{1}\) or \(P_{2}\) with equal probability. (ii) A blocky configuration that tiles the lattice with \(20\times20\) squares of a single strategy, producing straight, flat interfaces at \(t=0\). (iii) The same random mix as in (i) but augmented with a permanent, uniform mutation rate \(\mu=10^{-6}\). For each \(\beta\) we ran \(30\) independent trajectories of \(10^{5}\) MCS and recorded the mean fraction of every strategy together with one standard deviation.

\begin{figure}[t]
	\centering
	\includegraphics[width=0.95\textwidth]{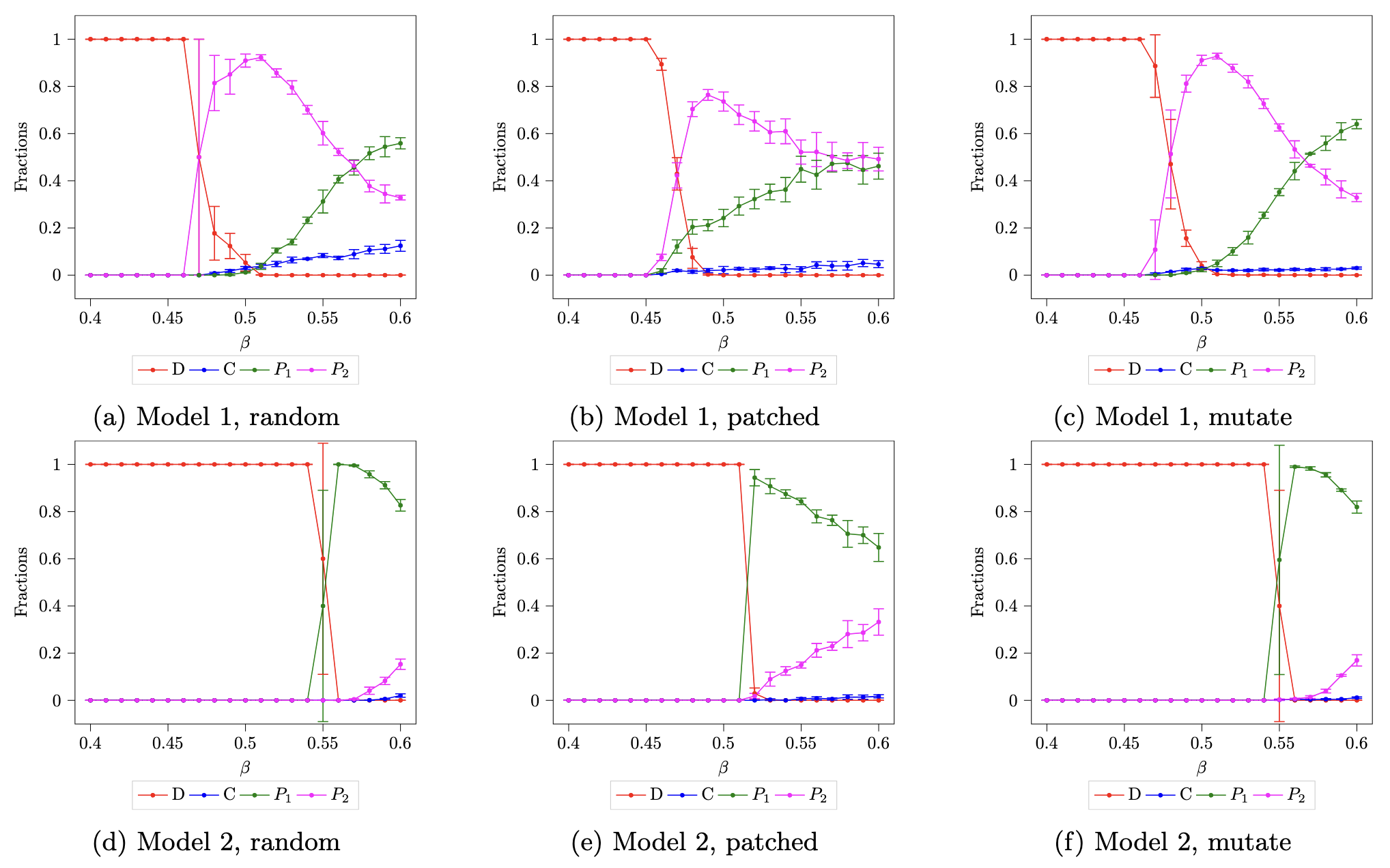}
\caption{Stationary strategy fractions as a function of the fine $\beta$ from 0.4 to 0.6 at fixed $r=2.5$ and $\gamma=0.4$. Every symbol reports the mean of thirty realizations on a $1200\times1200$ lattice evolved for $10^{5}$ MCS; vertical bars mark one standard deviation. Top row: Model~1 under (a) random initialization, (b) $20\times20$ patchy blocks, and (c) random start with mutation rate $\mu=10^{-6}$. Bottom row: the same protocols for Model~2. All initialization methods consistently confirm the critical fine advantage of Model 1 ($\beta_c \approx 0.46$) over Model 2 ($\beta_c \approx 0.54$), demonstrating that the efficiency gain is a robust dynamical property independent of initial spatial organization. The concordance between random initialization and weak mutation protocols validates our methodological approach for mapping equilibrium phase boundaries, while the patchy initialization reveals path-dependent effects that quantitatively shift but do not eliminate the relative ranking between punishment strategies.}

	\label{fig:initialization}
\end{figure}

Figure~\ref{fig:initialization} distills the outcomes of the largest simulation set into six comparative panels, each summarizing the cumulative dynamics of our long-run runs. The upper row reports Model~1, the lower row Model~2. In every panel the curves exhibit a sharp kink that identifies the critical fine. The kink sits at \(\beta_{c}\approx0.46\) for Model~1 and \(\beta_{c}\approx0.54\) for Model~2 (except the patched initialization). Hence the relative efficiency advantage of the norm-responsive design is a genuine dynamical property.

Although every simulation converges to a well-defined stationary mix of strategies, the fine value at which defectors lose global dominance is not universal. With fully random initial conditions (Figure~\ref{fig:initialization}, panels (a) and (d)) the decisive kink in the order parameters appears at \(\beta_{c}\approx0.46\) for Model~1 but only at \(\beta_{c}\approx0.54\) for Model~2. The broader window required by Model 2 reflects the additional resources wasted when \(P_{2}\) players escalate punishment inside defector strongholds instead of at cooperative frontiers. When the lattice is seeded with $20\times20$ homogeneous blocks (Figure~\ref{fig:initialization}, panels (b) and (e)) the early-time turbulence is suppressed and the competing domains meet along perfectly flat fronts. In this artificially clean geometry the threshold in Model~2 shifts downward to \(\beta_{c}\approx0.51\), confirming that a portion of the gap between the two designs originates from stochastic interface roughening rather than from deterministic payoff differences alone. Remarkably, despite these quantitative shifts, the relative ranking between Model~1 and Model~2 remains invariant across all initialization schemes, confirming the robustness of the context-sensitive advantage.

The weak-mutation runs (Figure~\ref{fig:initialization}, panels (c) and (f)) tell a decisive methodological story. Because every lattice site has a probability \(\mu = 10^{-6}\) of spontaneously switching strategy at each step, the system is continually nudged out of accidental local optima without ever disturbing the payoff hierarchy. The trajectories evolved with weak mutation equilibrate swiftly and, most importantly, converge to exactly the same stationary fractions and critical fines as those obtained from fully random initializations. This concordance demonstrates that a minute mutation rate supplies sufficient exploration to escape metastable traps while leaving the genuine phase structure unaltered. In other words, a vanishingly small trickle of mutations is sufficient to restore the ergodicity that patchy seeds break, but it does not alter the underlying phase structure. We note that this technique is specially effective when the characteristic length scale is large which makes proper system behavior unattainable at small sizes~\cite{helbing_pre10,lee2024suppressing,shen_cp25}.

We therefore rely on random initialization and random initialization with weak mutation to accelerate the sweep across the vast \((\beta,\gamma)\) grid, while validating every phase boundary with extensive random-start runs. The fact that both protocols reproduce \(\beta_{c}^{(1)}\approx0.46\) and \(\beta_{c}^{(2)}\approx0.54\) to within statistical error confirms that the 15\% efficiency gap separating the two punishment designs is a genuine dynamical property, not a numerical artifact.

\subsection{Microscopic Mechanisms: Domain Formation and Invasion}

\begin{figure}[h]
	\centering
	\includegraphics[width=1\textwidth]{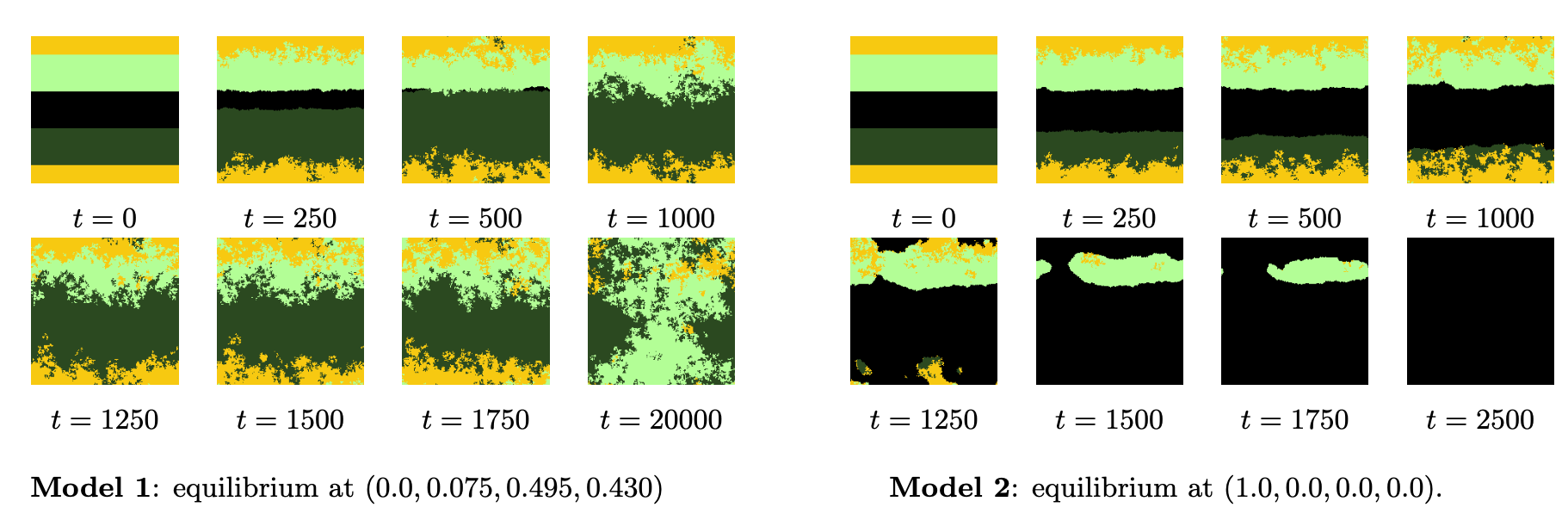}
\caption{Spatiotemporal evolution of the strategy distribution under the two punishment designs. Simulations start from a striped initial condition in which horizontal bands of equal width are populated respectively with defectors, cooperators, conventional punishers, and norm-responsive punishers, giving the overall composition $(D,C,P_{1},P_{2}) = (0.25,0.25,0.25,0.25)$ on a $200 \times 200$ lattice. The parameter set is $r = 2.5$, $\beta = 0.5$, $\gamma = 0.4$. The left column displays successive snapshots for Model~1, where norm-responsive punishers intensify sanctions when local cooperation is high; the right column presents the matching snapshots for Model~2, in which the same punishers intensify sanctions when cooperation is low. Color coding: black $=$ defectors ($D$), yellow $=$ cooperators ($C$), light-green $=$ conventional punishers ($P_{1}$), dark-green $=$ norm-responsive punishers ($P_{2}$). The sequence reveals the fundamental mechanism behind Model 1's efficiency: norm-responsive punishers (dark green) form protective buffers along cooperative fronts and propagate as self-reinforcing waves, while in Model 2 they waste resources in defector-dominated regions and are quickly eliminated, demonstrating why context-sensitive targeting of cooperative-defector interfaces generates superior invasion dynamics.}

	\label{fig:snapshots1}
\end{figure}

\begin{figure}[h]
	\centering
	\includegraphics[width=1\textwidth]{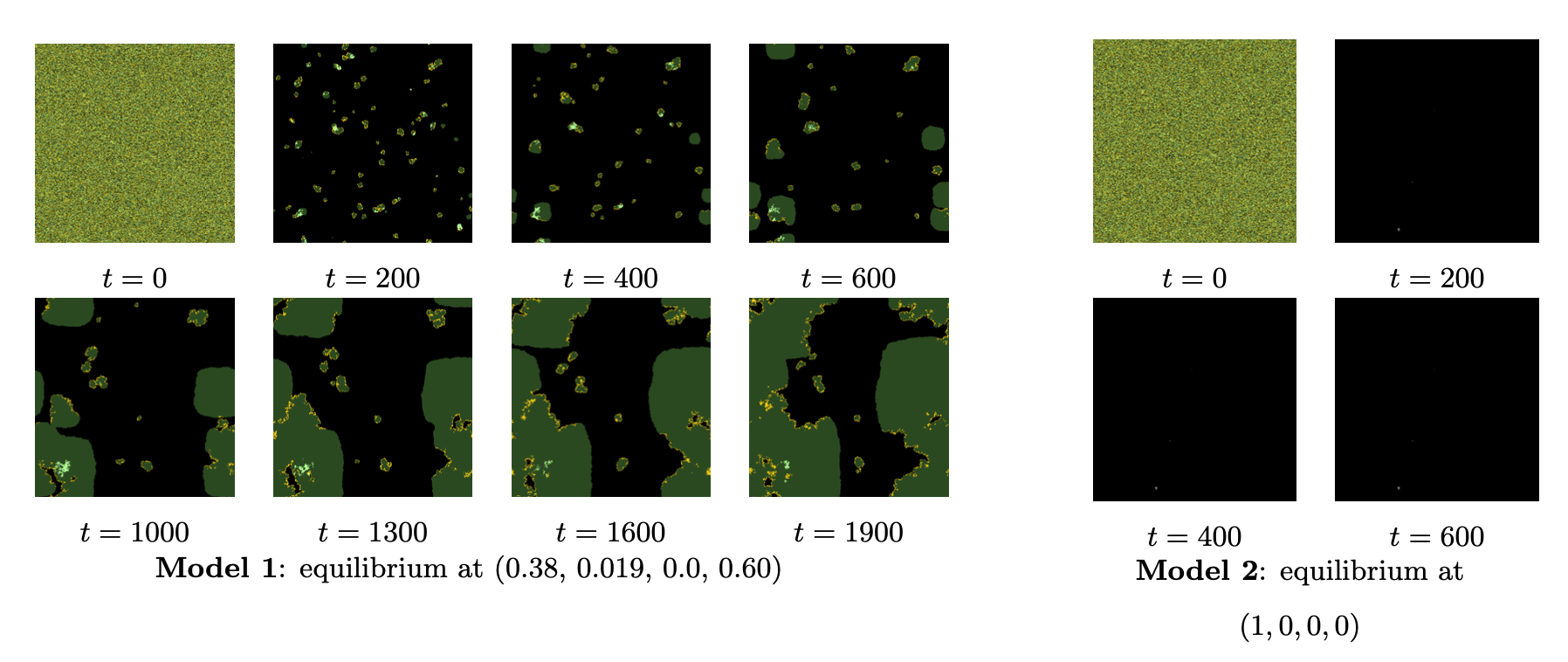}
\caption{Spatiotemporal evolution of the strategy distribution under the two punishment designs. Simulations commence from a fully random configuration in which each lattice site is assigned $D$, $C$, $P_{1}$, or $P_{2}$ with equal probability, yielding the initial composition $(D,C,P_{1},P_{2}) = (0.25,0.25,0.25,0.25)$ on a $200 \times 200$ lattice. The parameter set is $r = 2.5$, $\beta = 0.5$, $\gamma = 0.4$. The left column shows successive snapshots for Model~1, where norm-responsive punishers intensify sanctions when local cooperation is high; the right column presents the corresponding snapshots for Model~2, where intensification occurs in low-cooperation neighborhoods. The color code is identical to used in Fig.~\ref{fig:snapshots1}. Starting from maximal disorder, Model 1 demonstrates superior cluster formation and expansion: cooperative patches (yellow) survive and grow protected by expanding dark-green halos of norm-responsive punishers, creating a labyrinthine network that eventually dominates the lattice, while Model 2 shows rapid collapse of isolated punishment islands in the hostile defector sea, illustrating how resource conservation in unfavorable environments enables more effective deployment at strategic boundaries.}

	\label{fig:snapshots2}
\end{figure}

\begin{figure}[h!]
	\centering
	\includegraphics[width=1\textwidth]{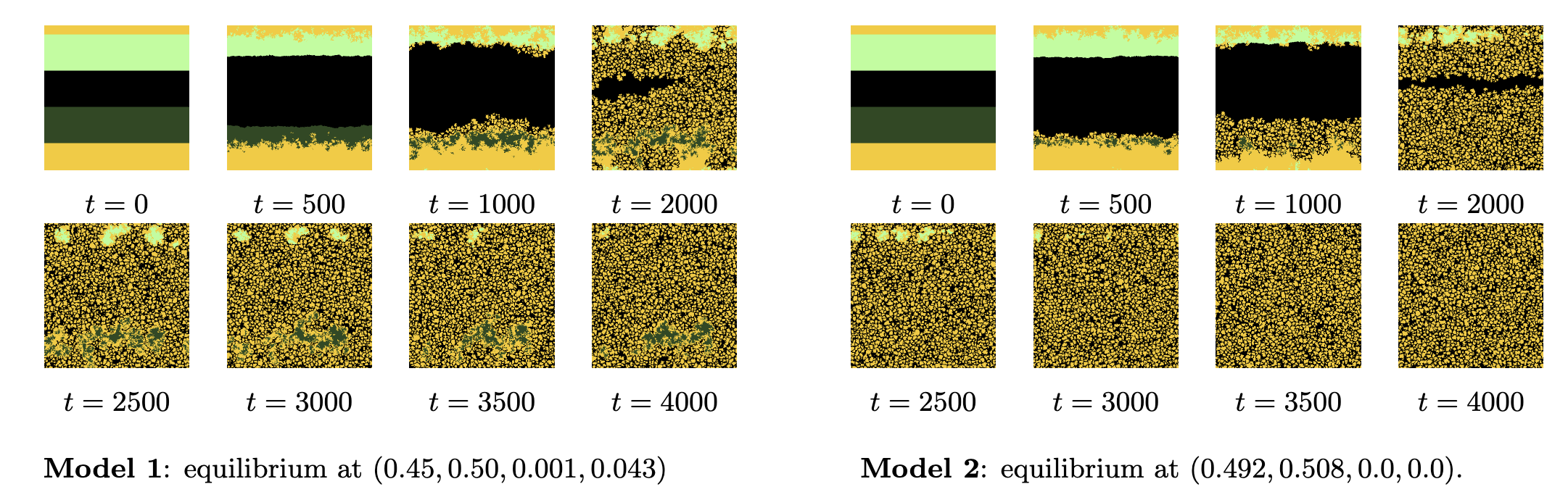}
\caption{Spatiotemporal evolution of the strategy distribution under the two punishment designs. Simulations start from a striped initial condition in which horizontal bands of equal width are populated respectively with defectors, cooperators, conventional punishers, and norm-responsive punishers, giving the overall composition $(D,C,P_{1},P_{2}) = (0.25,0.25,0.25,0.25)$ on a $200 \times 200$ lattice. The parameter set is $r = 4.0$, $\beta = 0.5$, $\gamma = 0.9$. The left column displays successive snapshots for Model~1, where norm-responsive punishers intensify sanctions when local cooperation is high; the right column presents the matching snapshots for Model~2, in which the same punishers intensify sanctions when cooperation is low. The color code is identical to used in Fig.~\ref{fig:snapshots1}. At high synergy factor $r=4.0$, network reciprocity already supports substantial cooperation (yellow terraces), yet Model 1 still maintains strategic advantage by using norm-responsive punishers as boundary guardians that cement cooperative zones, while Model 2's misallocated enforcement in defector territory leads to punisher extinction and subsequent cooperative erosion, demonstrating that context-sensitive targeting remains beneficial even when baseline cooperation is strong.}

	\label{fig:snapshots3}
\end{figure}

\begin{figure}[h!]
	\centering
	\includegraphics[width=1\textwidth]{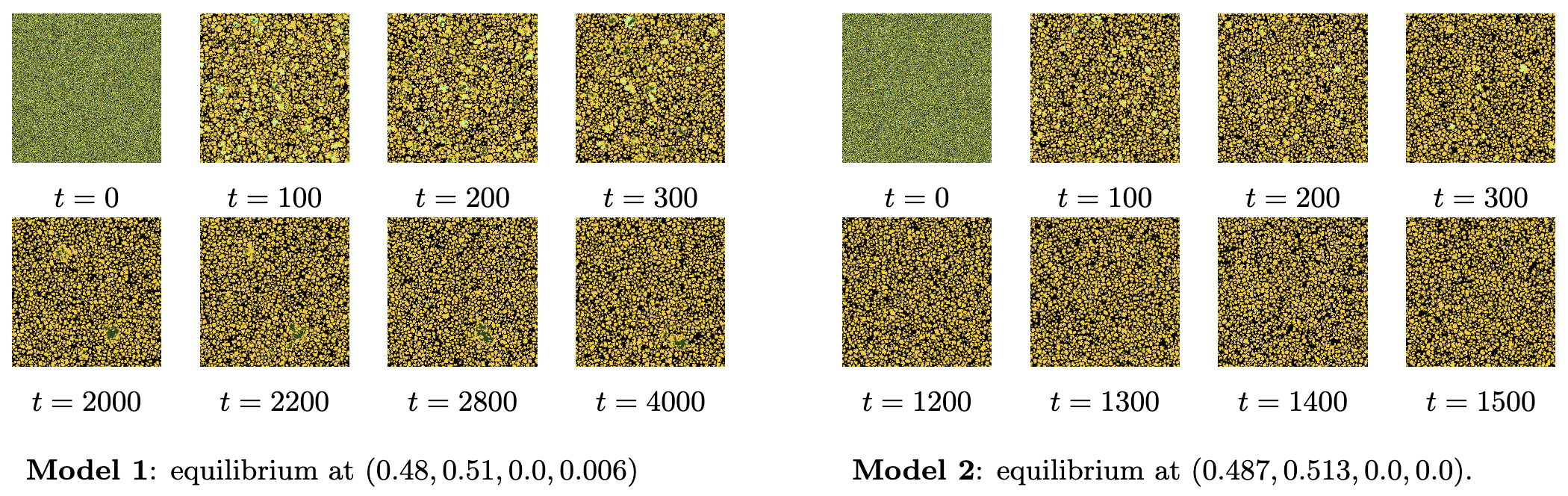}
\caption{Spatiotemporal evolution of the strategy distribution under the two punishment designs. Simulations commence from a fully random configuration in which each lattice site is assigned $D$, $C$, $P_{1}$, or $P_{2}$ with equal probability, yielding the initial composition $(D,C,P_{1},P_{2}) = (0.25,0.25,0.25,0.25)$ on a $200 \times 200$ lattice. The parameter set is $r = 4.0$, $\beta = 0.5$, $\gamma = 0.9$. The left column shows successive snapshots for Model~1, where norm-responsive punishers intensify sanctions when local cooperation is high; the right column presents the corresponding snapshots for Model~2, where intensification occurs in low-cooperation neighborhoods. The color code is identical to used in Fig.~\ref{fig:snapshots1}. Under high synergy conditions with random initialization, rapid cooperative clustering occurs naturally (dense yellow networks), but Model 1 preserves roaming enforcement islands that eliminate residual defectors, while Model 2 loses all norm-responsive punishers and settles into a vulnerable pure cooperation state with scattered defector specks, showing that even when cooperation is strongly favored, strategic punishment allocation provides insurance against exploitation and maintains system resilience.}

	\label{fig:snapshots4}
\end{figure}

Direct visualization of spatiotemporal evolution reveals the mechanistic origin of context-sensitive punishment's efficiency advantage through real-time tracking of domain boundaries and invasion dynamics. Figures~\ref{fig:snapshots1}–\ref{fig:snapshots4} record the evolution in real time and reveal the local encounters that shape the macroscopic phase diagrams. Two qualitatively different synergy factors are considered, $r=2.5$ and $r=4.0$. For each $r$ we perform one run from a striped initialization, which sets up four straight interfaces at $t=0$, and a second run from a fully random seed, which imposes maximal initial disorder. The striped initialization proves crucial for isolating the pure effects of interface dynamics by eliminating stochastic fluctuations that can obscure the fundamental invasion mechanisms, while the random initialization demonstrates robustness under realistic conditions. The punishment parameters are fixed within each synergy class, namely $(\beta,\gamma)=(0.5,0.4)$ for $r=2.5$ and $(\beta,\gamma)=(0.5,0.9)$ for $r=4.0$. The same color legend is used throughout: black denotes defectors $D$, yellow cooperators $C$, light green conventional punishers $P_{1}$ and dark green norm-responsive punishers $P_{2}$.

\paragraph*{Striped start, $r = 2.5$ (Fig.~\ref{fig:snapshots1})}

The perfectly flat interfaces created by striped initialization provide an ideal laboratory for studying the fundamental physics of cooperative invasion without confounding geometric irregularities. At $t = 0$ the lattice consists of four horizontal stripes, hence four perfectly flat interfaces. In Model~1 (left column) a site at the front switches the norm-responsive punishers to their high-intensity setting as soon as the local cooperation fraction exceeds $1/2$. The boosted $P_{2}$ players immediately etch a narrow dark-green buffer that protects the adjacent cooperators and, at the same time, invades the neighboring defectors. The invasion front propagates until the lattice freezes into the mixed state $\bigl(f_{D},\,f_{C},\,f_{P_{1}},\,f_{P_{2}}\bigr)\simeq(0,\,0.075,\,0.495,\,0.430)$. In Model~2 (right column) escalation is triggered deep inside the black defector band, a region where punishment is expensive and conversion is unlikely. The $P_{2}$ layer therefore burns out, $P_{1}$ cannot stem the counter-invasion, and global defection is established well before $t = 3,000$ MCS.

\paragraph*{Random start, $r = 2.5$ (Fig.~\ref{fig:snapshots2})}
With fully disordered initial conditions the early dynamics is a coarsening race among countless microscopic clusters. In Model~1 the yellow cooperative speckles survive long enough for $P_{2}$ to activate along their perimeters; the expanding rims fuse into a labyrinthine network that eventually suffocates the black background. A thin halo of cooperators remains because an isolated $C$ completely surrounded by $P_{2}$ neighbors is immune to exploitation. Model~2 again spends most of its fine in territories where punishment yields no strategic benefit; dark-green islands are isolated in a hostile sea and vanish quickly, leaving only shrinking droplets of $P_{1}$. The contrasting invasion histories explain why the critical fine required to eliminate defectors is $\beta_{c}\simeq0.46$ for Model~1 but $\beta_{c}\simeq0.54$ for Model~2, as quantified in Fig.~\ref{fig:beta_scan}.

\paragraph*{Striped start, $r = 4.0$ (Fig.~\ref{fig:snapshots3})}
At this high synergy the public good already supports a stable background of cooperators, so the invasion game now unfolds on a landscape where yellow terraces are present from the outset. In Model~1 (left column) the norm-responsive punishers can no longer rely on carving new territory; instead they cement the existing $C$ zones into a equilibrium $\bigl(f_{D},\,f_{C},\,f_{P_{1}},\,f_{P_{2}}\bigr)\approx(0.49,\,0.50,\,0.001,\,0.0043)$. Here opportunistic cooperators harvest the public good while a thin veneer of $P_{2}$ controls the boundaries and keeps defectors in check. Model~2 (right column) squanders its limited budget where it yields no return. The dark-green $P_{2}$ line fires off costly escalations while still buried in black territory and vanishes first; the light-green $P_{1}$ band, still exposed, is swiftly overrun by the advancing defectors. Once the green buffer collapses the black stripe sweeps laterally across the lattice, eroding the yellow plateau until only scattered cooperative islets survive in a dominant sea of defectors. The stationary state is therefore a black field peppered with small yellow patches and entirely devoid of punishers, illustrating that escalation launched from inside hostile territory is an evolutionary dead end at high synergy.

\paragraph*{Random start, $r = 4.0$ (Fig.~\ref{fig:snapshots4})}
With fully random initial conditions the large value of $r$ induces rapid clustering: within a few hundred 
MCSs the lattice is dominated by a dense yellow mesh of cooperators. Model~1 preserves a faint sprinkling of $P_{2}$ islands that roam the network and annihilate residual defectors, whereas Model~2 loses the norm-responsive type entirely and settles into an almost pure $C$ phase whose serenity is disrupted only by rare black specks. The qualitative contrast between the sanctioning schemes survives the change in initial disorder, although the quantitative gap in the critical fine narrows because strong baseline reciprocity already favors cooperation when $r = 4.0$.

These spatiotemporal analyses converge on a fundamental principle: optimal punishment strategies must target the dynamic interfaces between cooperation and defection, where marginal enforcement investments yield maximum behavioral conversion rates. Model~1 internalizes this lesson. Its $P_{2}$ players remain quiescent inside defector strongholds, conserve their budget, then unleash doubled fines exactly where a few cooperators have already softened the target. Figures~\ref{fig:snapshots1} and \ref{fig:snapshots2} show the dark-green wave advancing in lockstep with the expanding yellow frontier, while Figures~\ref{fig:snapshots3} and \ref{fig:snapshots4} reveal that the same tactic still pays when synergy alone is strong enough to seed wide cooperative terraces. Model~2 spends the same amount of resources but in the wrong place. By escalating deep inside hostile black domains it burns capital without tipping the local balance; the squandered effort allows conventional punishers, or defectors themselves, to reclaim the terrain. These contrasting invasion films supply the missing causal chain that links the microscopic rules to the macroscopic phase diagrams: context-aware sanctioning reallocates an unchanged punishment budget to the handful of lattice sites where each fine buys the greatest behavioral change. The payoff is a consistent reduction of approximately 15\% in the critical fine required to secure a cooperative society.

\clearpage

\section{Conclusion}
\label{sec:conclusions}

Our study shows that an enforcement policy attuned to micro-scale social cues can enhance efficiency while reducing costs of cooperation while preserving its moral force. When sanction intensity is conditioned on the immediate social environment, even a modest rule that doubles both fine and cost only when the focal group is already at least half cooperative can extinguish defection at noticeably lower statutory fine levels than a rule that escalates indiscriminately. The resulting efficiency dividend, roughly 15\% across all cost levels and synergy factors we explored, appears even though both punisher types incur the same marginal expense per act. This efficiency gain reveals that spatial evolutionary dynamics contain hidden optimization opportunities that can be exploited through context-sensitive institutional design. This finding exposes a reservoir of latent efficiency that is unlocked by matching the incentive to the micro-state of the population. It also suggests a broader theoretical point: behavioral feedbacks that align costly enforcement with moments of heightened prosocial conformity can leverage existing pockets of cooperation, turning them into self-propagating fronts that spread through the system. Such context-aware mechanisms could help reconcile the twin policy goals of minimizing enforcement budgets while maximizing compliance, and they provide a conceptual bridge between classic models of peer punishment and emerging frameworks that emphasize adaptive, data-driven regulation. This approach complements recent advances in punishment theory, including graded punishment mechanisms~\cite{quan2023cooperation} and state-dependent incentive allocation protocols~\cite{sun2023state}, while offering a novel perspective on how punishment intensity can be optimally distributed across heterogeneous social environments. Future empirical and theoretical work can enlarge this bridge by testing alternative escalation profiles, allowing the cooperation threshold itself to evolve, or embedding the mechanism in richer social networks and institutional architectures.

Monitoring the time evolution of spatial patterns starting from different initial state helped us to reveal main mechanism which makes context-sensitive norm so effective. In Model~1 the norm-responsive punishers first allow a thin yellow beachhead of cooperators to crystallize at the edge of a defector domain; only when that local majority tips beyond $50\%$ do they unleash the doubled sanction. The freshly converted nodes immediately reinforce the majority condition in neighboring groups, so the intensified fines sweep laterally along the front rather than dissipating inward. Every black site that flips widens the cooperative strip, amplifying the majority signal, and the domino continues until a coherent dark-green wall overtakes the lattice. This autocatalytic process creates a positive feedback loop where successful conversions locally amplify enforcement intensity, generating self-sustaining waves of cooperative behavior that propagate through the system. Model~2 burns the same fiscal fuel in a very different furnace. Here the escalated fine is triggered wherever cooperators are scarce, so most of the budget is spent deep inside black deserts where the probability of successful conversion is vanishing. The few defectors who do switch side fall back into minority groups on the next update and are quickly re-absorbed. Punishers exhaust their resources without generating a self-sustaining front, and the remnant green specks are wiped out by the advancing black tide. The microscopic contrast translates directly into the macroscopic gap in the critical fine: one design turns finite resources into a traveling wave of compliance, the other disperses them into thermal noise. More broadly the videos teach a transferable lesson. Enforcement achieves its greatest leverage when it targets actors perched at the tipping-point between violation and conformity, because every successful conversion creates new majority pockets that propagate the incentive on their own. Investing in hopeless cases or guaranteed saints merely dilutes the impact; focusing on the undecided can ignite a chain reaction that spreads cooperation at a fraction of the cost. These invasion dynamics align with broader insights from spatial evolutionary games showing how topological features and local clustering fundamentally shape cooperation outcomes~\cite{szolnoki2009topology, brandt2003punishment}, and demonstrate the critical importance of targeting interventions at strategic boundary locations.

These findings translate into three actionable principles for institutional design. First, penalty schedules should be conditional on measurable compliance indicators rather than fixed by statute~\cite{szolnoki_jtb13}. Second, inspectors should concentrate on the fault lines that separate mostly compliant from largely non-compliant regions, because a single intervention there can shift the balance of an entire neighborhood. Third, flooding chronically non-compliant areas with sanctions may waste resources that would do more good elsewhere. Implementing these context-sensitive strategies could yield enforcement cost reductions of 15\% while maintaining equivalent compliance levels, representing substantial efficiency gains for resource-constrained institutions. These efficiency gains align with previous demonstrations that small fractions of strategically designed agents can generate substantial welfare improvements~\cite{szolnoki2020blocking, lee2021small, zhang2022adaptive}, suggesting that targeted interventions consistently outperform blanket approaches across diverse social dilemma contexts. The practical relevance of these principles is further supported by empirical studies showing that punishment institutions can be effectively selected and sustained through voting and learning mechanisms~\cite{vasconcelos2022punishment}, indicating that context-sensitive enforcement strategies may be both theoretically optimal and socially acceptable.

Future research should prioritize three extensions: optimizing the cooperation threshold through evolutionary dynamics, developing continuous rather than binary escalation functions, and testing robustness across diverse network topologies including scale-free and multilayer architectures. Given the growing interest in dynamic and multi-player trust games with sophisticated punishment mechanisms~\cite{wang2025dynamic}, our framework could be extended to incorporate multiple cooperation thresholds, temporal punishment delays, and heterogeneous agent capabilities. Allowing punishers to coordinate, signal, or learn could also reveal richer collective dynamics and new forms of strategic adaptation. Beyond the specific setting of public goods, the principle uncovered here speaks to a broad class of social interventions. Whether the aim is to curb tax evasion, reduce littering, or promote vaccination, the most powerful lever is often not the severity of the sanction but the precision with which it is applied. This precision-based approach represents a natural evolution from earlier work on adaptive mechanisms in evolutionary games and smart behavioral strategies \cite{lee2024suppressing}, pointing toward a new generation of context-aware institutional designs. A context-aware policy uses the same budget to greater effect, aligning enforcement pressure with the locations where social tipping points are within reach. As demonstrated by recent work on institutional incentives~\cite{flores2024evolution} and information sharing mechanisms~\cite{szolnoki2013information}, the integration of contextual awareness into policy design represents a promising avenue for enhancing collective welfare while minimizing institutional costs. By aligning enforcement intensity with local social dynamics, institutions can achieve the dual objectives of maximizing compliance and minimizing costs, providing a scientifically grounded framework for building resilient cooperative societies in an era of limited resources.

\section*{Data and Code Availability}
The complete source code for the Monte Carlo simulations, including parameter sweeps and visualization routines, is available at \url{https://github.com/artermi/Norm_Enforcement_2024}. The code is implemented in C$++$ and includes detailed documentation for reproducibility.

\vspace{0.5cm}
This research was supported by the National Research, Development and Innovation Office (NKFIH) under Grant No. K142948.

\bibliographystyle{elsarticle-num-names}
\bibliography{ref}

\end{document}